\documentclass{iopart}
\usepackage{graphicx}
\usepackage[colorlinks=true,citecolor=blue,urlcolor=blue]{hyperref}
\usepackage{xcolor}
\usepackage{orcidlink}
\usepackage{siunitx}

\usepackage[mathb]{mathabx} 

\newcommand{\rtHz}{$\sqrt{\rm Hz}$}
\newcommand{\Finesse}{\mathcal{F}}

\begin{document}
\begin{flushleft}
{\rm RESCEU-3/25}
\end{flushleft}

\title[Initial acquisition requirements for DECIGO and B-DECIGO]{Initial acquisition requirements for optical cavities\\ in the space gravitational wave antennae DECIGO and B-DECIGO}

\author{Yuta~Michimura$^{1,2,\star}$\orcidlink{0000-0002-2218-4002}, Koji~Nagano$^{3,4,\dagger}$\orcidlink{0000-0001-6686-1637}, Kentaro~Komori$^{1,5}$\orcidlink{0000-0002-4092-9602}, Kiwamu~Izumi$^6$\orcidlink{0000-0003-3405-8334}, Takahiro~Ito$^6$\orcidlink{0000-0003-1491-1940}, Satoshi~Ikari$^7$\orcidlink{0000-0002-0467-768X}, Tomotada~Akutsu$^8$\orcidlink{0000-0003-0733-7530}, Masaki~Ando$^{5,1}$\orcidlink{0000-0002-8865-9998}, Isao~Kawano$^6$, Mitsuru~Musha$^9$\orcidlink{0000-0001-7741-4584}, Shuichi~Sato$^{10}$\orcidlink{0000-0001-5560-5224}}

\address{$^1$ Research Center for the Early Universe (RESCEU), Graduate School of Science, University of Tokyo, Bunkyo, Tokyo 113-0033, Japan}
\address{$^2$ Kavli Institute for the Physics and Mathematics of the Universe (Kavli IPMU), WPI, UTIAS, University of Tokyo, Kashiwa, Chiba 277-8568, Japan}
\address{$^3$ Institute for Multidisciplinary Sciences, Yokohama National University, Yokohama, Kanagawa 240-8501, Japan}
\address{$^4$ LQUOM, Inc., Yokohama, Kanagawa 240-8501, Japan}
\address{$^5$ Department of Physics, The University of Tokyo, Bunkyo, Tokyo 113-0033, Japan}
\address{$^6$ Institute of Space and Astronautical Science, Japan Aerospace Exploration Agency, Sagamihara, Kanagawa 252-5210, Japan}
\address{$^7$ Department of Aeronautics and Astronautics, University of Tokyo, Bunkyo, Tokyo 113-8656, Japan}
\address{$^8$ National Astronomical Observatory of Japan, Mitaka, Tokyo 181-8588, Japan}
\address{$^9$ Institute for Laser Science, University of Electro-Communications, Chofu, Tokyo 182-8585, Japan}
\address{$^{10}$ Department of Science and Engineering, Hosei University, Koganei, Tokyo 184-8584, Japan}
\ead{$^{\star}$michimura@resceu.s.u-tokyo.ac.jp, $^{\dagger}$nagano-koji-hv@ynu.ac.jp}
\vspace{10pt}
\begin{indented}
\item[]\today
\end{indented}

\begin{abstract}
DECIGO (DECi-hertz Interferometer Gravitational Wave Observatory) is a space-based gravitational wave antenna concept targeting the 0.1--10 Hz band. It consists of three spacecraft arranged in an equilateral triangle with 1,000 km sides, forming Fabry-P{\'e}rot cavities between them. A precursor mission, B-DECIGO, is also planned, featuring a smaller 100 km triangle. Operating these cavities requires ultra-precise formation flying, where inter-mirror distance and alignment must be precisely controlled. Achieving this necessitates a sequential improvement in precision using various sensors and actuators, from the deployment of the spacecraft to laser link acquisition and ultimately to the control of the Fabry-P{\'e}rot cavities to maintain resonance. In this paper, we derive the precision requirements at each stage and discuss the feasibility of achieving them. We show that the relative speed between cavity mirrors must be controlled at the sub-micrometer-per-second level and that relative alignment must be maintained at the sub-microradian level to obtain control signals from the Fabry-P{\'e}rot cavities of DECIGO and B-DECIGO.
\end{abstract}

\maketitle

\section{Introduction}
Since the groundbreaking direct detection in 2015, the ground-based laser interferometric gravitational wave observatory network, consisting of LIGO~\cite{aLIGO}, Virgo~\cite{AdV}, and KAGRA~\cite{AsoKAGRA,PTEP01KAGRA}, has reported the detection of more than 250 gravitational wave events. These ground-based interferometers are sensitive to the audioband frequency range from approximately 10 Hz to 10 kHz, making them well-suited for detecting binary mergers involving stellar-mass black holes and neutron stars~\cite{ObservingScenarioPaper}. Next-generation detectors such as the Einstein Telescope~\cite{ET} and Cosmic Explorer~\cite{CE}, which aim to be an order of magnitude larger, are currently underway. Once operational, these detectors are expected to detect nearly all stellar-mass binary mergers in the observable universe.

In the 2030s, the space-based observatory LISA~\cite{LISA2017,LISAPathfinder}, sensitive to the millihertz frequency range, will enable the observation of massive black hole binaries. Similar projects, including TianQin~\cite{TianQin2016,TianQin2021} and Taiji~\cite{Taiji2017,Taiji2020}, are also under development. At even lower frequencies in the nanohertz range, a global network of pulsar timing arrays has recently reported strong evidence of the stochastic gravitational wave background~\cite{NanoGrav,EPTAInPTA,PPTA,CPTA} and is expected to make further progress in the coming years.

As gravitational wave observations expand across different frequency bands, the frequency range of 0.1 Hz to 10 Hz has become increasingly important as a new frontier~\cite{AMIGO,DO}. This decihertz band is particularly well suited for detecting intermediate-mass black hole binaries~\cite{Matsubayashi2004} and probing primordial gravitational wave background originating from cosmic inflation~\cite{Nakayama2008,Calcagni2021}. Additionally, it provides an early warning of stellar-mass binary mergers~\cite{Liu2020} and offers insights into their astrophysical origins~\cite{DECIGO-smoking-gun}.

However, achieving high-sensitivity observations in this frequency range requires significant technological developments. Ground-based gravitational wave detectors suffer from limitations due to seismic noise and Newtonian gravity gradient noise, making observations in this band challenging~\cite{MANGO,RanaRMP}. While various concepts have been proposed for space-based optical transponder-type laser interferometers similar to LISA~\cite{AMIGO}, shifting sensitivity from the millihertz range to the decihertz range requires shortening the arm length so that it remains smaller than the wavelength of gravitational waves to prevent signal cancellation. Studies of ALIA~\cite{ALIA} and BBO~\cite{BBO} have shown that this, in turn, necessitates high-power laser sources on the order of several hundred watts to reduce shot noise, posing significant technical challenges. Another proposed approach, TianGO~\cite{TianGO}, aims to construct a Michelson interferometer in space and reduce shot noise using quantum squeezing techniques. Furthermore, there is the Lunar Gravitational-Wave Antenna mission proposal, which aims to utilize the Moon as a gravitational wave detector to explore the decihertz range~\cite{LGWA2021,LGWA2024}.

Among various space-based proposals in the decihertz range, DECIGO (DECi-hertz Interferometer Gravitational wave Observatory) represents a distinct approach by forming Fabry-P{\'e}rot cavities between spacecraft to reduce shot noise~\cite{DECIGO2001,Kawamura2019,DECIGO2021}. Fabry-P{\'e}rot cavities have been used in ground-based gravitational wave observatories for many years. Utilizing them as a high-sensitivity gravitational wave detector requires suppressing mirror displacement fluctuations to well below the typical laser wavelength of $1~\si{\micro m}$ and controlling mirror angular motions to the nanoradian level~\cite{AsoKAGRA,aLIGOASC,Michimura2017}. Achieving this level of stability necessitates high-precision formation flying of spacecraft, which fundamentally differs from the constellation flying employed by LISA, where an optical transponder-type laser interferometer allows for variations in the distances between spacecraft. Such high-precision formation flying has not yet been demonstrated in orbit~\cite{Ito2024,SILVIA}. Therefore, an incremental approach will be required for the initial acquisition, incorporating multiple sensors, such as Global Navigation Satellite System (GNSS)-based relative navigation and laser ranging techniques demonstrated by the GRACE Follow-On mission~\cite{GRACE-FOLaserRanging}, to progressively achieve the precision necessary to operate the Fabry-P{\'e}rot cavities.

In this paper, we divide the initial acquisition sequence into three stages: deployment of the three spacecraft, laser link acquisition, and lock acquisition to maintain cavity resonance. We derive the precision requirements for each stage and discuss the sensors and actuators necessary to achieve them. Section~\ref{sec:Configuration} provides a brief overview of the interferometer design of DECIGO and its precursor mission B-DECIGO. Section~\ref{sec:InitialAcquisition} defines each stage of the initial acquisition sequence, while Section~\ref{sec:Requirements} discusses the corresponding precision requirements. Finally, our conclusions are summarized in Section~\ref{sec:Conclusions}.

\section{Interferometer design} \label{sec:Configuration}

\begin{figure}[t]
    \centering
    \includegraphics[width=0.8\textwidth]{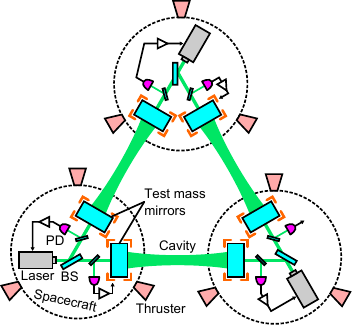}
    \caption{Schematic view of DECIGO and B-DECIGO. For simplicity, components such as electro-optic modulators, steering mirrors, and mode-matching telescopes are not included in the diagram. The dual-pass differential Fabry-P{\'e}rot interferometer scheme is shown as an example interferometer control scheme. BS: beam splitter, PD: photodiode.}
    \label{fig:concept}
\end{figure}

\begin{figure}[t]
    \centering
    \includegraphics[width=0.6\textwidth]{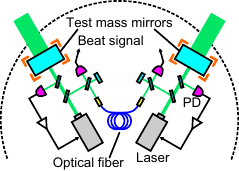}
    \caption{Schematic view of the optical setup in one spacecraft for the back-linked Fabry-P{\'e}rot interferometer control scheme.}
    \label{fig:BLFPI}
\end{figure}

\subsection{Configuration and control scheme} \label{sec:scheme}

The conceptual design of DECIGO and B-DECIGO is shown in Fig.~\ref{fig:concept}. A single cluster consists of three spacecraft forming an equilateral triangle with a side length of 1,000 km for DECIGO and 100 km for B-DECIGO. Each spacecraft houses two cavity mirrors that act as free-falling test masses above the control bandwidth. These mirrors form Fabry-P{\'e}rot cavities between the spacecraft, with laser beams injected from both sides of each cavity, making them dual-pass cavities. Arm length fluctuations are measured from the cavity reflected beams, while gravitational wave signals are extracted from the differential arm length signals. One cluster of three spacecraft effectively functions as three interferometric gravitational wave detectors.

Each spacecraft is equipped with thrusters for drag-free control. Each mirror has local sensors to measure its relative displacement and orientation with respect to the spacecraft. By using signals from both the local sensors and the Fabry-P{\'e}rot cavities, the thrusters control the position and attitude of the spacecraft, achieving drag-free control similar to what has been demonstrated with LISA Pathfinder~\cite{LISAPathfinder}. As a result, the spacecraft follows the free-falling test mass mirrors.

There are multiple ways to obtain the differential arm length signals and to control the interferometers. One approach is to recombine the reflected light from the Fabry-P{\'e}rot cavities to form a Fabry-P{\'e}rot Michelson interferometer, similar to the ground-based detectors. This method has the advantage of optically removing common-mode noise such as laser frequency and intensity noises. However, displacement noise from the beam splitter could contaminate the gravitational wave readout, and it also requires additional control of the differential Michelson length and alignment.

To avoid these complexities, two methods, namely dual-pass differential Fabry-P{\'e}rot interferometer (DPDFPI) scheme~\cite{NaganoThesis,Nagano2021} and back-linked Fabry-P{\'e}rot interferometer (BLFPI) scheme~\cite{Izumi2021,Sugimoto2024}, have been recently proposed and demonstrated through table-top experiments. In the DPDFPI scheme, each spacecraft has a single laser source, and the incident beam is split into two arms by a beam splitter. Five out of six length error signals, obtained from the cavity reflection beams detected by photodiodes, are fed back to either the test mass actuators or the laser frequency actuators in order to maintain the resonant conditions of the cavities (Fig.~\ref{fig:concept}). One length error signal cannot be fed back to these length or frequency degrees of freedom, as one of them must serve as a reference. Therefore, to keep the error signals within their linear range, the absolute lengths of the three arms need to be adjusted~\cite{Nagano2021}. While this scheme offers a simple configuration, it imposes additional requirements on the initial acquisition sequence.

In the BLFPI scheme, on the other hand, each spacecraft has two independent laser sources, with each laser frequency stabilized to the length of its respective arm cavity (Fig.~\ref{fig:BLFPI}). The differential arm length signal is then obtained from the heterodyne beat signals. This approach eliminates the need for continuous feedback of the length signals to the test mass mirror actuators, which introduces actuator noise~\cite{Michimura2017}, but instead requires a low noise frequency discriminator to read out the beat signal~\cite{Izumi2021}. Since the initial acquisition sequence imposes more stringent requirements on the DPDFPI scheme, we assume DPDFPI scheme for the remainder of this paper.

In both cases, to obtain length and angular error signals from the arm cavities, the Pound-Drever-Hall (PDH) method~\cite{PDH} and the wavefront sensing (WFS) method~\cite{WFS} are expected to be used. Both methods employ phase modulation of the incident light using an electro-optic modulator. The PDH method extracts cavity length fluctuations from the phase difference between the resonant carrier and the non-resonant sidebands, while WFS extracts angular fluctuations of cavity mirrors from the phase difference between the carrier and sideband wavefronts.

\subsection{Interferometer parameters}
\begin{table}[t]
\begin{center}
  \caption{The fiducial interferometer parameters of DECIGO and B-DECIGO used for calculating the requirements of the initial acquisition sequence~\cite{DECIGO2021,NaganoThesis}. The symbols used to represent them are also shown. The strain sensitivities listed here indicate the requirements at 0.1~Hz with a safety factor of 10. Critically-coupled symmetric cavities are assumed. BS: beam splitter, RoC: radius of curvature, RIN: relative intensity noise.} \label{table:parameters}
\begin{tabular}{lcccc}
   \br
   & & DECIGO & B-DECIGO \\
   \hline
   Strain sensitivity (1/\rtHz) & $h_{\rm req}$ & $1 \times 10^{-24}$ & $1 \times 10^{-23}$ \\
   Arm length (km) & $L$ & 1,000 & 100 \\
   Laser wavelength (nm) & $\lambda$ & 515 & 515 \\
   Power at BS (W) & $P_0$ & 10 & 0.02 \\
   Finesse & $\Finesse$ & 10 & 100 \\
   RoC of mirror (km) & $R$ & 607 & 60.7 \\
   Beam radius at mirror (m) & $w$ & 0.46 & 0.15 \\
   Mirror mass (kg) & $m$  & 100 & 30 \\
   Mirror radius (m) & $r$ & 0.83 & 0.35 \\
   Mirror thickness (m) & $t$ & 0.02 & 0.04 \\
   \multicolumn{3}{l}{Laser noise requirements at 0.1~Hz}  & \\
   \, Frequency noise (Hz/\rtHz) & $\delta \nu$ & 1 & 1 \\
   \, RIN (1/\rtHz) & $I_{\rm RIN}$ & $2 \times 10^{-9}$ & $1 \times 10^{-8}$ \\
   \multicolumn{3}{l}{Frequency actuator requirements}  & \\
   \, Range (MHz) & $\Delta \nu_{\rm act}$ & 200 & 200 \\
   \multicolumn{3}{l}{Mirror actuator requirements}  & \\
   \, Range (\si{\micro N}) & $F_{\rm max}$ & 2 & 6 \\
   \br
\end{tabular}
\end{center}
\end{table}

The interferometer parameters of DECIGO and B-DECIGO are summarized in Table~\ref{table:parameters}. Since the sensitivity is currently being redesigned~\cite{Ishikawa2021,Iwaguchi2021,Kawasaki2022,Ishikawa2023}, these parameters serve as fiducial values for calculating the requirements for the initial acquisition sequence in this paper.

The strain sensitivity requirement, mirror mass, laser power, arm cavity length, and finesse for DECIGO are determined to enable the observation of primordial gravitational waves~\cite{DECIGO2021}. B-DECIGO, while having an arm length that is ten times shorter, features a higher finesse to maintain the same cavity pole frequency as DECIGO. Here, the cavity pole is the half width at half maximum of the cavity and is given by $f_{\rm c}=c/(4 L \Finesse)$, where $L$ and $\Finesse$ are the length and the finesse of the cavity, and is about 7.5 Hz for both DECIGO and B-DECIGO. A laser wavelength of 515~nm was chosen because a shorter wavelength results in a smaller beam size and it allows the use of iodine absorption lines for frequency stabilization~\cite{Suemasa2019}. The laser noise requirements are determined to achieve the strain sensitivity requirement~\cite{NaganoThesis}.

To derive the mirror sizes, we assumed fused silica density mirrors and an aspect ratio of $2r/t=5$, where $r$ and $t$ are mirror radius and thickness, respectively. This aspect ratio was chosen to keep optical losses sufficiently low, but further optimization, including thermal noise considerations, is a subject for future investigation~\cite{Iwaguchi2021,Kawasaki2022}. Using a mirror $g$-factor of $g=1-L/R=-0.65$ to ensure a stable cavity and tolerable optical angular anti-spring effect~\cite{SiggSidles,Hirose2010}, the radius of curvature of the mirrors and the beam radius at the mirror are derived, assuming a symmetric cavity. With these parameters, the diffraction loss due to the finite mirror size is 0.33\% for DECIGO and 23~ppm for B-DECIGO, which is small enough to achieve the required arm cavity finesse.

\subsection{Actuator requirements}
For the requirement on the range of frequency actuator, we have used typical values for laser sources and acousto-optic modulators. The required force range of the mirror actuators is determined to be sufficient to compensate for the forces from laser radiation pressure and Earth's gravity while remaining small enough to minimize the injection of actuator noise.

In the case of DECIGO, which will be placed in a heliocentric orbit, laser radiation pressure becomes the dominant force. Since the intracavity power is given by $P_{\rm cav}=P_0 \Finesse/\pi$, the force from laser radiation pressure can be calculated with
\begin{equation}
  F_{\rm cav} = \frac{2 P_{\rm cav}}{c} = \frac{2 \Finesse P_0}{\pi c},
\end{equation}
where $c$ is the speed of light. For DECIGO parameters, $F_{\rm cav}= 0.2$~\si{\micro N}, and the actuator range of $F_{\rm max}= 2$~\si{\micro N} should be sufficient. For B-DECIGO paramerers, $F_{\rm cav}= 4$~\si{n N}, and is negligible compared with the Earth's gravity. Note that $P_0$ represents the power incident at the beam splitter (``BS" in Fig.~\ref{fig:concept}). After splitting, the power incident into each cavity in one direction is $P_0/2$. Since laser beams are injected from both ends of the arm cavity, the total intracavity power $P_{\rm cav}$ defined here corresponds to the power that would build up in the cavity if $P_0$ were injected from a single side.

For B-DECIGO, the orbit has not yet been determined, but the primary candidate currently under consideration is a geocentric orbit. In this case, the contribution of Earth's gravity will be more than an order of magnitude larger. The force perturbations due to Earth's $J_2$ gravity potential are given by~\cite{Alfriend2010,Ito2024},
\begin{equation}
  F_{\rm J_2} \simeq \frac{3 J_2 G M_{\Earth} R_{\Earth}^2 m L}{2(R_{\Earth}+H)^5}.
\end{equation}
Here, $J_2=1.08 \times 10^{-3}$ is the $J_2$ coefficient, $G$ is gravitational constant, $H$ is altitude, and $M_{\Earth}$ and $R_{\Earth}$ are Earth mass and radius, respectively. For the geosynchronous equatorial orbit, $H=3.6 \times 10^{7}~\si{m}$, this force perturbations will be $F_{\rm J_2} = 0.6~\si{\micro N}$, and the actuator range of $F_{\rm max}= 6~\si{\micro N}$ should be sufficient.

The force noise requirement for each mirror, derived from the strain sensitivity requirement $h_{\rm req}$, is given by
\begin{equation}
  \delta F_{\rm req} = \frac{m h_{\rm req} L}{2 (2 \pi f_{\rm req})^2} .
\end{equation}
Here, $f_{\rm req}=0.1$~\si{Hz} is the frequency at which the requirement is defined, and we assumed that the force noises acting on the four mirrors are uncorrelated. For DECIGO and B-DECIGO, this requirement is calculated to be $2 \times 10^{-17}$~N/\rtHz\ and $6 \times 10^{-18}$~N/\rtHz, respectively. The development of actuators with both such low noise and large dynamic range remains an important subject for future work, though not addressed in detail in this paper.

\section{Initial acquisition sequence} \label{sec:InitialAcquisition}

\begin{figure}[t]
    \centering
    \includegraphics[width=0.9\textwidth]{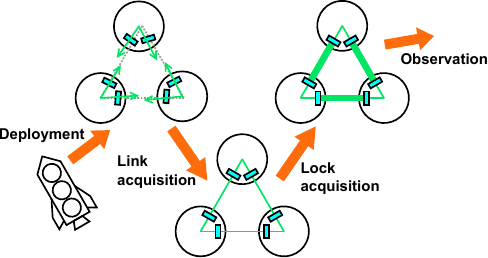}
    \caption{Schemetic of the initial acquisition sequence.}
    \label{fig:sequence}
\end{figure}

The three spacecraft of DECIGO and B-DECIGO, once launched into orbit, must go through three stages before gravitational wave observations can begin: the spacecraft deployment stage, the inter-spacecraft laser link acquisition stage, and the lock acquisition stage to maintain the resonant condition of the optical cavities. These three stages are collectively referred to as the initial sequence in this paper, and the conceptual diagram is shown in Fig.~\ref{fig:sequence}.

The goal of the deployment stage is to position the three spacecraft in an equilateral triangle with the required precision in the difference of absolute distances between them. Additionally, the goal is to align the spacecraft attitudes to a level where link acquisition can begin and to set the beam pointing for the laser interferometer.

The goal of the link acquisition stage is to align the beam pointing direction and mirror orientation so that light can occasionally resonate, or flash, and to intermittently apply control, or lock, to maintain the resonant condition of the Fabry-P{\'e}rot cavities. Here, flashing occurs because the cavity length is not yet controlled, and resonance happens only when the round-trip cavity length coincidentally matches an integer multiple of the laser wavelength. In order to confirm flashing, this process requires reducing the relative velocity between the mirrors using signals from the laser beam and other relevant sensors.

The goal of the lock acquisition stage is to establish and maintain a continuous locked state of the Fabry-P{\'e}rot cavities. This involves further reducing the relative velocity between the mirrors and suppressing arm length fluctuations and mirror alignment fluctuations to meet the sensitivity requirements.

The requirement calculations described in the next section for each stage do not assume specific sensors or actuators. For reference, Table~\ref{table:sensors-actuators} summarizes examples of sensors and actuators that could be used at each stage. It is important to note that spacecraft sensors such as GNSS and star trackers measure absolute position and attitude of the spacecraft, whereas Fabry-P{\'e}rot cavities provide relative measurements, including inter-mirror distance fluctuations and alignment fluctuations between the input beam axis and the cavity axis defined by the alignment of the test mass mirrors.

\begin{table}[t]
\begin{center}
  \caption{Examples of sensors and actuators that could be used at each stage of the initial sequence. For distance and angle sensors, those that measure the absolute distance between spacecraft and those that measure changes in distance are shown. Steering mirrors are used to steer the input beam axis into the arm cavity. GNSS: Global Navigation Satellite System, AOD: acousto-optic deflector~\cite{Musha2021}, QPD: quadrant photodiode, PDH: Pound-Drever-Hall~\cite{PDH}, WFS: wavefront sensor~\cite{WFS}.} \label{table:sensors-actuators}
\small
\begin{tabular}{lccc}
   \br
   & Deployment & Link acquisition & Lock acquisition \\
   \hline
   Absolute position sensor    & GNSS & GNSS & GNSS \\
   Absolute attitude sensor & Star tracker, & Star tracker, & Star tracker, \\
                            & Gyroscope & Gyroscope & Gyroscope \\
   Longitudinal distance    &  AOD (absolute) & AOD (absolute), & Arm cavity \\
   sensor                   &                 & Laser ranging (relative) & PDH (relative)  \\
   Angle \& lateral distance &           & AOD (absolute), & Arm cavity \\
   sensor (two-dimensional)   &                 & QPD (relative) & WFS (relative) \\
   Position actuator          & Thrusters & Thrusters & Thrusters, \\
                              &           &           & laser frequency, \\
                              &           &           & test mass mirrors \\
   Alignment actuator        & Thrusters & Thrusters,         & Thrusters, \\
                             &            & steering mirrors, & steering mirrors, \\
                             &            & test mass mirrors & test mass mirrors \\
    \br
\end{tabular}
\end{center}
\end{table}

\section{Requirements} \label{sec:Requirements}
In this section, we derive the precision requirements for the initial sequence of DECIGO and B-DECIGO at each stage and discuss the feasibility to achieve these requirements. Table~\ref{table:requirements} summarizes the derived requirements. Throughout this section, the velocity requirement refers specifically to the relative velocity between the cavity mirrors along the cavity axis, since this component is the most relevant for the initial acquisition process.

\subsection{Deployment}
For laser interferometric gravitational wave detectors, it is not necessarily crucial to achieve the absolute arm length exactly as designed. However, minimizing the difference between arm lengths is important from a noise reduction perspective. Since laser interferometers are not inherently designed to precisely match absolute lengths, we consider achieving the required precision in arm length differences using spacecraft sensors from the deployment stage.

The requirements for the arm length difference depend on the interferometer configuration. Here, we focus on the case of the DPDFPI scheme, where the requirements are more stringent than the other schemes described in Sec.~\ref{sec:scheme}. As discussed, the error signal that is not used for the length control must be sufficiently close to the cavity resonance point. If it is detuned, laser intensity noise coupling will affect the sensitivity. From this intensity noise coupling, the requirement for matching the lengths of the equilateral triangle is given by~\cite{Nagano2021}
\begin{equation}
  \Delta L < \frac{2 L^2 h_{\rm req}}{\lambda I_{\rm RIN}} .
\end{equation}
Here, $I_{\rm RIN}$ denotes the relative intensity noise, defined as the laser intensity fluctuations normalized by its average value. Note that this requirement is not derived from laser frequency noise, because the coupling of laser frequency noise does not depend on the detuning.

As summarized in Table~\ref{table:requirements}, this is 2~km for DECIGO and 40~m for B-DECIGO. These values are sufficiently feasible, since the accuracy of position measurements using GNSS can reach sub-meter level~\cite{PRISMA2012}. GNSS cannot be used at distances farther than the geosynchronous equatorial orbit, but alternative approaches, such as absolute ranging techniques developed for LISA~\cite{Yamakoh2024} or the acousto-optic deflector (AOD)-based method~\cite{Musha2021}, could be employed. For LISA, a method using the time-of-flight of pseudorandom noise has been developed, and sub-meter ranging accuracy has been demonstrated on-ground~\cite{Yamakoh2024}. In the AOD approach, the deflection angle of the beam depends on the applied modulation frequency, and the absolute cavity length can be determined by measuring the microwave phase of the beat between the deflected beams. A preliminary test reported in Ref.~\cite{Musha2021} demonstrated a resolution of 6.25 m, indicating that meeting our requirement is feasible. Both methods are autonomous and do not require continuous ground intervention.

It is worth noting that these requirements are essential for the simultaneous operation of all three gravitational wave detectors to achieve the desired sensitivity. If only two detectors are required to be operational, meeting this requirement is not strictly necessary. Also, this requirement is stricter than the typical requirement for ground-based detectors, where achieving $\Delta L / L$ smaller than 1\% is sufficient to achieve a factor of 100 in common-mode rejection of laser noises.

Additionally, the spacecraft attitudes need to be adjusted to a certain level during the deployment stage to point the laser beam within the range that can be scanned during the link acquisition phase. Here, we set the laser beam pointing requirement to be the same as the GRACE Follow-On mission's link acquisition scanning range, which is $\Delta \theta_{\rm beam} < 3$~mrad~\cite{GRACE-FO-laser-link}. This is larger than the typical accuracy of the attitude sensors of spacecraft, such as star tracker. However, due to the mounting accuracy of the optical components relative to the spacecraft sensors, there is a possibility that an unknown offset could be introduced into the laser beam axis. Therefore, the 3~mrad requirement can be considered as a requirement for the mounting accuracy of the optical components. This requirement applies not only to the in-plane angle (yaw) of the equilateral triangle but also to the out-of-plane angle (pitch), as is the case for other angular requirements discussed in the following sections.

\subsection{Link acquisition}
In order for the arm cavities to begin flashing, the motion of the beam spot on the test mass mirrors must be kept smaller than the beam radius. Therefore, the required precision for the pointing direction of the interferometer beam is given by
\begin{equation}
  \Delta \theta_{\rm beam} < \frac{w}{L} .
\end{equation}
Similarly, the required precision for the alignment of the test mass mirrors is given by
\begin{equation}
  \Delta \theta < \frac{w}{2L} .
\end{equation}
Here, the factor of two arises from the fact that a change in the mirror alignment by an angle $\theta$ causes the reflected beam to shift by $2\theta$. As summarized in Table~\ref{table:requirements}, these requirements will be on the order of microradians to sub-microradians. These precision levels are well within the capabilities of beam position sensors, such as quadrant photodiodes or position sensitive detectors. In DECIGO and B-DECIGO, the mirror radius is about 1.8 and 2.3 times larger than the beam radius, respectively. A beam displacement of one beam radius $w$ from the mirror center therefore causes additional diffraction loss. However, the resulting loss is estimated to be less than 20\% and 5\%, respectively, and is acceptable for the link acquisition process.

To confirm the cavity flashing, the relative velocity between the mirrors must be sufficiently reduced so that a sufficient amount of laser power can build up inside the cavity. Given that the storage time of the cavity $\tau = 2 L \Finesse/(\pi c)$ must be smaller than the time it takes to pass through the cavity linewidth $t_{\rm lw}=\lambda/(2 \Finesse v)$, the requirement on the relative velocity between the cavity mirrors is given by~\cite{Rakhmanov2001}
\begin{equation} \label{eq:lock-start-velocity}
 v < \frac{\pi c \lambda}{4 L \Finesse^2} .
\end{equation}
This requirement is 1.2~\si{\micro m/s} for DECIGO and 0.12~\si{\micro m/s} for B-DECIGO, which is significantly more stringent than the velocity achievable with current satellite sensors. For example, the value achieved in the PRISMA mission using GPS-based navigation was on the order of 1~mm/s~\cite{PRISMA2012}.

Therefore, additional sensors are required to measure and further reduce the velocity to complete the link acquisition stage. To achieve this, methods using acousto-optic deflectors~\cite{Musha2021}, optical frequency combs~\cite{Shen2022}, and laser ranging techniques demonstrated by the GRACE Follow-On mission~\cite{GRACE-FOLaserRanging} are being considered. Alternatively, velocity sensing could be achieved by counting interference fringe changes per unit time or using time-of-flight measurements based on laser intensity modulation at radio frequencies. These methods are planned to be demonstrated in orbit as part of the SILVIA (Space Interferometer Laboratory Voyaging towards Innovative Applications) mission, which is currently under development with a target launch in the early 2030s~\cite{SILVIA}.

\begin{table}[t]
 \begin{center}
  \caption{Summary of requirements for the initial sequence of DECIGO and B-DECIGO. The values listed here represent the maximum allowable limits calculated using the parameters in Table~\ref{table:parameters}, and all actual values are required to be smaller than these limits. Angluar requirements apply to both in pitch and yaw. ``Same" indicates that the requirement is the same as in the previous stage.} \label{table:requirements}
\small
\begin{tabular}{lccc}
  \br
  DECIGO & Deployment & Link acquisition & Lock acquisition\\
         & goal & goal & goal\\
  \hline
  Cavity length difference $\Delta L$    & 2~km & same & same \\
  Cavity length fluctuations $\delta L_{\rm rms}$ & - & 34~cm & 0.5~nm \\
  Relative velocity between mirrors $v$     & - & 1.2~$\si{\micro m/s}$ & 23~nm/s \\
  Beam pointing $\Delta \theta_{\rm beam}$ & 3~mrad & 0.46~$\si{\micro rad}$ & - \\
  Mirror relative alignment $\Delta \theta$ & - & 0.23~$\si{\micro rad}$ & 0.035~nrad \\
  \hline
  B-DECIGO & & & \\
  \hline
  Cavity length difference $\Delta L$    & 40~m & same & same \\
  Cavity length fluctuations $\delta L_{\rm rms}$ & - & 3.4~cm & 0.1~nm \\
  Relative velocity between mirrors $v$     & - & 0.12~$\si{\micro m/s}$ & 23~nm/s \\
  Beam pointing $\Delta \theta_{\rm beam}$ & 3~mrad & 1.5~$\si{\micro rad}$ & - \\
  Mirror relative alignment $\Delta \theta$ & - & 0.73~$\si{\micro rad}$ & 0.35~nrad \\
  \br
\end{tabular}
\end{center}
\end{table}

\subsection{Lock acquisition}
Let us first examine whether the velocity requirement given by Eq.~(\ref{eq:lock-start-velocity}) is sufficient to initiate the acquisition of cavity lock. If only mirror actuators are used for lock acquisition, the force must be applied within the time it takes to pass through the cavity linewidth $t_{\rm lw}$ to suppress the mirror momentum $mv$~\cite{Michimura2017}. Since this impulse $mv/t_{\rm lw}$ must be smaller than the actuator force range $F_{\rm max}$, the requirement is given by
\begin{equation}
 v < \sqrt{\frac{\lambda F_{\rm max}}{2 \Finesse m}} .
\end{equation}
For both DECIGO and B-DECIGO, this value is 23~\si{nm/s}, which is much more stringent than the requirement given by Eq.~(\ref{eq:lock-start-velocity}), indicating that lock acquisition using only the mirror actuators would be challenging. This requirement can be achieved once the lock is acquired.

Therefore, the use of laser frequency actuators, such as piezoelectric actuators of the laser source and acousto-optic modulators, is considered for the lock acquisition. If the cavity length fluctuation before locking is too large, it would exceed the range of the frequency actuators. Hence, the cavity length fluctuation must satisfy
\begin{equation} \label{eq:lock-start-rms}
 \delta L_{\rm rms} < L \frac{\Delta \nu_{\rm act}}{\nu} ,
\end{equation}
where $\nu=c/\lambda$ is the laser frequency. This requirement is 34~\si{cm} for DECIGO and 3.4~\si{cm} for B-DECIGO, which is within the precision reach of GNSS-based navigation in the low Earth orbit, as demonstrated by the PRISMA mission~\cite{PRISMA2012}. While achieving such precision in the orbits considered for DECIGO and B-DECIGO is currently challenging, it is expected to be realized by using additional sensors capable of \si{\micro m/s} level measurements, as discussed in the previous section. Therefore, we set this as an additional requirement for the link acquisition stage, as summarized in Table~\ref{table:requirements}.

Since maintaining a locked state for an extended period using only frequency control is difficult due to potential drifts in arm length and laser frequency, decelerating the mirrors using the feedback signal from the frequency control must begin immediately after lock acquisition. Therefore, a requirement for the relative velocity between the mirrors $v$ must also be imposed to maintain the lock. Using the time required to start mirror deceleration control after lock acquisition $\Delta t_{\rm TM}$, the requirement is given by
\begin{equation}
 v < \frac{\delta L_{\rm rms}}{\Delta t_{\rm TM}} = \frac{L \Delta \nu_{\rm act}}{\nu \Delta t_{\rm TM}} .
\end{equation}
When $\Delta t_{\rm TM}=10$~\si{s}, chosen to account for potential delays in signal processing, this requirement is 3.4~\si{cm/s} for DECIGO and 0.34~\si{cm/s} for B-DECIGO. These are less stringent than the requirement given by Eq.~(\ref{eq:lock-start-velocity}).

Additionally, to decelerate the mirrors using the feedback signal of the frequency control, the deceleration control must be applied within the duration $t_{\rm fl} = L \Delta \nu_{\rm act}/(v \nu)$ of the frequency lock. Since the impulse $mv/t_{\rm fl}$ must be smaller than the actuator force range $F_{\rm max}$, another requirement on the velocity is set by
\begin{equation}
 v < \sqrt{\frac{L F_{\rm max} \Delta \nu_{\rm act}}{m \nu}} .
\end{equation}
This requirement is 83~\si{\micro m/s} for both DECIGO and B-DECIGO, which is also less stringent than the requirement given by Eq.(\ref{eq:lock-start-velocity}). Therefore, once the velocity requirement in Eq.(\ref{eq:lock-start-velocity}) is satisfied, the lock acquisition step can be initiated.

The goal of the lock acquisition stage is to maintain the cavity resonance to a level where observation can begin. In order to sufficiently minimize laser intensity noise coupling, the residual cavity length fluctuation must satisfy
\begin{equation}
 \delta L_{\rm rms} < \frac{h_{\rm req} L}{I_{\rm RIN}} .
\end{equation}
This value is 0.5~\si{nm} for DECIGO and 0.1~\si{nm} for B-DECIGO. Usually, intensity noise coupling is the dominant first-order effect, while other perturbative contributions, such as thermal effects and sensor and actuator nonlinearities, will be the subject of future work. Additionally, the required control gain to stabilize the original laser frequency noise $\delta \nu$ using the arm cavity length must meet
\begin{equation}
 G_{\rm freq} > \frac{\delta \nu}{h_{\rm req} \nu} .
\end{equation}
This value is $1.7 \times 10^9$ for DECIGO and $1.7 \times 10^8$ for B-DECIGO. If post-processing is used to subtract frequency noise, the control gain can be lower than this value.

\begin{figure}[t]
    \centering
    \includegraphics[width=0.6\textwidth]{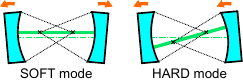}
    \caption{SOFT and HARD modes of angular motion of the cavity mirrors in a Fabry-P{\'e}rot cavity. The SOFT mode induces a translation of the cavity axis, while the HARD mode induces a rotation of the cavity axis. Cross marks indicate the mirrors' centers of curvature.}
    \label{fig:SOFT-HARD}
\end{figure}

There are also requirements for residual angular fluctuations after mirror alignment control. If the beam spot is displaced by a distance $\Delta d$ from the center of the mirror, the angular fluctuation $\delta \theta$ will cause a cavity length fluctuation of $\Delta d \delta \theta$~\cite{AsoKAGRA,aLIGOASC}. Therefore,  $\Delta d \delta \theta < h_{\rm req} L$ must be satisfied to sufficiently minimize alignment noise coupling. The beam spot displacement $\Delta d$ is also caused by residual angular fluctuations of the mirrors. In Fabry-P{\'e}rot cavities, there exist two alignment fluctuations, namely SOFT mode and HARD mode~\cite{SiggSidles,Hirose2010}. As shown in Fig.~\ref{fig:SOFT-HARD}, the SOFT mode is an antisymmetric tilt of cavity mirrors that causes a translation of the cavity axis, where the radiation pressure acts to reduce the mechanical restoring torque. The HARD mode is an symmetric tilt of cavity mirrors that causes a rotation of the cavity axis, where the radiation pressure acts to enhance the mechanical restoring torque. For a symmetric cavity, SOFT and HARD alignment changes cause the beam spot displacement given by
\begin{eqnarray}
  \Delta d &=& R \Delta \theta^{\rm SOFT}, \\
  \Delta d &=& \frac{RL}{2R-L} \Delta \theta^{\rm HARD} .
\end{eqnarray}
By requiring $\Delta d_{\rm rms} < 0.1$~\si{mm}, which is a typical requirement for ground-based interferometers, the requirement on the mirror alignment fluctuation $\Delta \theta_{\rm rms}$ can be calculated. A more stringent requirement comes from the HARD mode, and the values are summarized in Table~\ref{table:requirements}. Although these values are unprecedented in an in-orbit environment, they are achievable by employing the PDH method~\cite{PDH} and the WFS method~\cite{WFS}, as commonly done in ground-based interferometers. A preliminary study on cavity length and alignment controls can be found in Ref.~\cite{NaganoThesis}. However, a comprehensive assessment in achievable mirror alignment stability is beyond the scope of this paper and remains an important subject for future work.

\section{Conclusions} \label{sec:Conclusions}
In this paper, we have summarized the requirements for the initial acquisition of DECIGO and B-DECIGO, from the deployment of the three spacecraft to the acquisition of cavity resonances. Our analysis shows that most of the requirements related to cavity length and alignment can be met using technologies that have already been demonstrated in previous experiments. However, we have identified an important task that the relative velocity between mirrors must be reduced from the millimeter-per-second scale to the micrometer-per-second scale in the period during the laser link acquisition stage to enable the start of lock acquisition. Additionally, the stability of the mirror alignment and laser beam pointing must be maintained at the sub-microradian level to initiate the lock acquisition. Since these requirements must be achieved without relying on signals from the Fabry-Pérot cavities, meeting them poses a significant challenge. Additional sensors, such as a Michelson interferometer, acousto-optic deflectors, or laser ranging techniques, could be employed to satisfy these requirements. Experimental validation of these techniques, along with an in-orbit demonstration of ultra-precise formation flying using these sensors, will play a key role in enabling future missions. By systematically summarizing the key requirements and assessing their feasibility with current technology, this work represents an important step toward realizing DECIGO and B-DECIGO.


\ack
This work was supported by JSPS KAKENHI Grant Nos.~20H05639, 20H05850, 20H05854, 23K13501, 24K00640, and 24K21546. KN is employed by LQUOM and receives financial support from the company.

\section*{References}
\bibliographystyle{iopart_num}
\bibliography{DECIGOInitialAcquisition}
\end{document}